\documentclass[twocolumn,showpacs,amsmath,amssymb,aps,pra,floatfix,nofootinbib,superscriptaddress]{revtex4-1}
\usepackage{graphicx,graphics,times,color}% Include figure files
\usepackage{bm}% bold math
\usepackage{longtable}
\usepackage{color}
\usepackage{lipsum}
\usepackage{pgfplots}
\pgfplotsset{width=7cm,compat=newest}
\usepgfplotslibrary{units}
\usepackage{tikz}
\usetikzlibrary{spy,backgrounds}
\usepackage{pgfplotstable}
\usepackage{subfigure}
\usepgfplotslibrary{patchplots,colormaps}
\usetikzlibrary{decorations}
\usepackage{array}
\newcolumntype{C}[1]{>{\centering\arraybackslash}p{#1}}

\usetikzlibrary{pgfplots.external}

\pgfplotsset{compat=newest}%
\tikzset{external/system call =
{latex -shell-escape -halt-on-error
-interaction=batchmode -jobname "\image"
"\texsource"; dvips -o "\image".ps
"\image".dvi}}
\pgfplotstableread{resultOfV30.txt}\vthirty
\pgfplotstableread{resultOfV100.txt}\vhundred
\pgfplotstableread{resultOfVm70.txt}\vmseventy
\pgfplotstableread{resultOfVm100.txt}\vmhundred
\def\be{\begin{equation}}
\def\ee{\end{equation}}
\def\ba{\begin{eqnarray}}
\def\ea{\end{eqnarray}}
\def\la{\langle}
\def\ra{\rangle}

\usepackage[version=3]{mhchem}

\graphicspath{{image/}}

\begin{document}

\title{Quantum Decoherence Timescales for Ionic Superposition States in Ion Channels}

\author{V. Salari, N. Moradi, F. Fazileh, and F. Shahbazi }
\affiliation{Department of Physics, Isfahan University of Technology, Isfahan 84156-83111, Iran}

\date{\today}

\begin{abstract}
There are many controversial and challenging discussions about quantum effects in microscopic structures in neurons of the human brain. The challenge is mainly because of quick decoherence of quantum states due to hot, wet and noisy environment of the brain which forbids long life coherence for brain processing. Despite these critical discussions, there are only a few number of published papers about numerical aspects of decoherence in neurons. Perhaps the most important issue is offered by Max Tegmark who has calculated decoherence times for the systems of "ions" and "microtubules" in neurons of the brain. In fact, Tegmark did not consider ion channels which are responsible for ions displacement through the membrane and are the building blocks of electrical membrane signals in the nervous system. Here, we would like to re-investigate decoherence times for ionic superposition states by using the data obtained via molecular dynamics simulations. Our main approach is according to what Tegmark has used before. In fact, Tegmark didn't consider the ion channel structure and his estimates are only simple approximations. In this paper, we focus on the small nano-scale part of KcsA ion channels which is called "selectivity filter" and has a key role in the operation of an ion channel. Our results for superposition states of potassium ions indicate that decoherence times are in the order of picoseconds which are 10-100 million times bigger than the order calculated by Tegmark. This decoherence time is still not enough for cognitive processing in the brain, however it can be adequate for quantum states of cooled ions in the filter to leave their quantum traces on the filter and action potentials.  \end{abstract}


\maketitle

\section{Introduction}
Mainstream cognitive neuroscience has far largely ignored the role of quantum physical effects in the neuronal processes 
underlying cognition and consciousness. Classical physics is viewed by most scientists today as an approximation to the more accurate quantum theory, and therefore due to the nature of this classical approximation the causal effects of our conscious activity on the material substrate may appear to be eliminated. 

Ion channels are proteins in the membrane of excitable cells (e.g. neurons) that cooperate for the onset and propagation of electrical signals across membranes by providing a highly selective conduction of charges bound to ions through a channel like structure. The question here is whether quantum superposition states in ion channels can affect signal propagation in nervous system of the brain? The challenge is mainly because of quick decoherence of quantum states due to hot, wet and noisy environment of the brain which forbids long life coherence for brain processing. Tegmark has calculated decoherence times for quantum superposition of ions crossing the entire membrane based on a simple ion-pore diffusion model \cite{Tegmark2}. The calculated decoherence times for crossing ions in his work are derived from the scattering with environmental particles based on the Coulomb interaction between ions and particles. He has assumed that ions are in a superposition state of "inside" and "outside" of the cell and are separated by a distance of 10 nm as the thickness of the membrane. In the view of atomic scaled resolution maps and recent molecular dynamics studies of the filter region in this protein, this type of interaction is oversimplified and the pore-diffusion scattering model is not applicable to describe ion protein interactions. Tegmark has shown that decoherence times are in the order of $10^{-19}s$ to $10^{-20}s$ which are not enough for quantum states to survive for a quantum processing.  In fact, Tegmark has used a simple model and didn't consider the real structure of an ion channel. On the other side, Hagan et al \cite{Hagan} and Rosa and Faber \cite{Rosa} have determined other estimations for decoherence times for quantum states of "microtubules" in neurons in which some of their results indicate that there are still possibilities for quantum coherent states to be effective in brain processing. A big picture of comparisons between the above approaches can be seen the Table 1, in which the main reason of decoherence is because of interactions between the system and environmental particles due to scattering. \\
Here, we would like to use similar approaches (e.g. Tegmark's calculations \cite{Tegmark2}) for the system of "ion channel" which is not studied before. In our approach, we use a rather more accurate model via using the real structure of an ion channel based on the data obtained from molecular dynamics simulation. 

\begin{table*}[t]\label{tab1}
\caption{The History of Calculations for Decoherence Times in Neurons}
\begin{ruledtabular}
\begin{tabular}{ccccc}
 System &Mechanism (Collision/Interaction)& Approach& Order of Decoherence time &Year\\
 \hline
Microtubule (MT)&Soliton- Ion &Tegmark& $10^{-13}s$ &2000~\cite{Tegmark2}\\
Neuron (ions) & Ion-Ion &Tegmark& $10^{-20}s$  & 2000~\cite{Tegmark2}\\
Neuron (ions) & Ion-Water &Tegmark& $10^{-20}s$  & 2000~\cite{Tegmark2}\\
Neuron (ions) & Ion-Distant Ions &Tegmark&  $10^{-19}s$ & 2000~\cite{Tegmark2}\\
Microtubule (MT)& Soliton- Ion&Hagan et al& $10^{-7}-10^{-6}s$ &2002~\cite{Hagan}\\
Microtubule (MT) & MT-Ion&Hagan et al&$10^{-2}-10^{-1}s$ &  2002~\cite{Hagan}\\
Microtubule (MT)& MT-Ion& Rosa and Faber &$10^{-9}s$ &  2004~\cite{Rosa}\\
Microtubule (MT)& MT-Dipole& Rosa and Faber &$10^{-16}s$ &  2004~\cite{Rosa}\\

\end{tabular}
\end{ruledtabular}
\raggedright \end{table*}

%However, to achieve this a proper encoding has to be designed to compromise the random output of the measurement.
\section{The Selectivity Filter Structure in Ion Channels}
The selectivity filter is a part of the protein forming a narrow tunnel inside the ion channel which is responsible for the selection process and fast conduction of ions across the membrane. The determination of atomic resolution structure of the ion channel and selectivity filter by Mac Kinnon led to the award 
of the Nobel prize for chemistry in 2003 \cite{Mackinnon}.  
   \begin{figure} \centering    
       \includegraphics[width=10.5cm,height=4.5cm,angle=0]{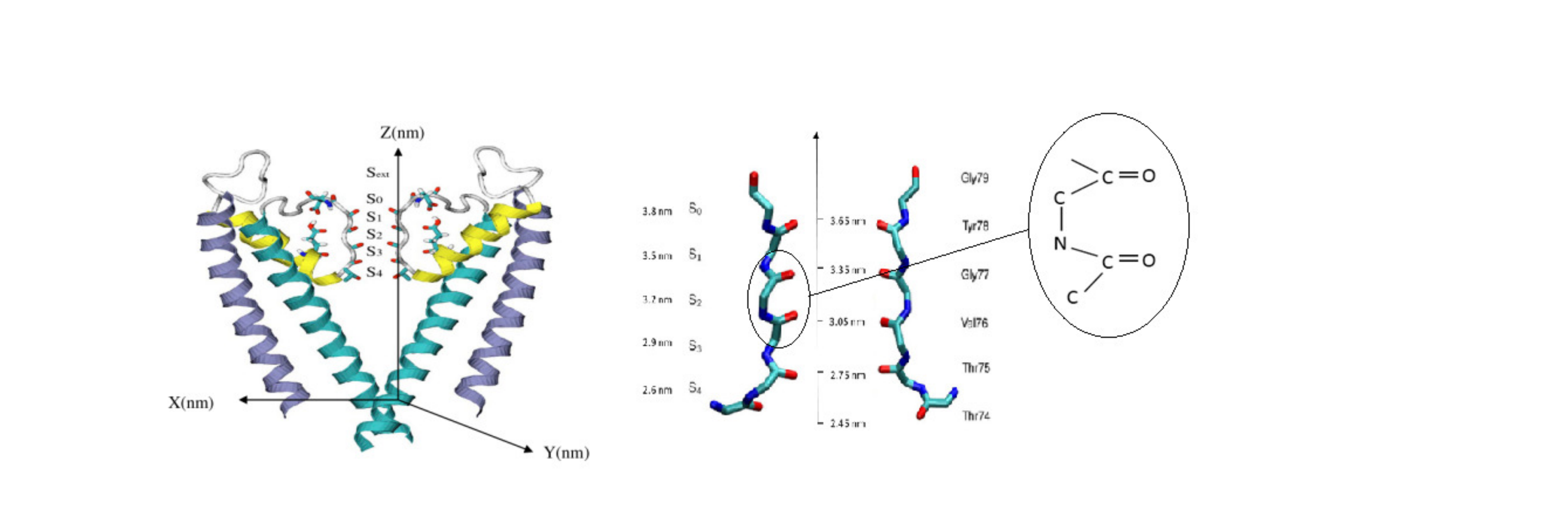}   
       \caption{ Left) A representation of $KcsA$ ion channel. Right) Two P-loop monomers in the selectivity filter, composed of the sequences of TVGYG amino acids $[T(Threonine, Thr75)$, $V(Valine, Val76) $, $G(Glycine, Gly77)$, $Y(Tyrosine, Tyr78)$, $G(Glycine, Gly79)]$ linked by peptide units H-N-C=O. }
        \label{fig2}   
   \end{figure}   
The $3.4$ nanometer long $KcsA$ channel is comprised of a $1.2$ nanometer long selectivity filter that is composed of four P-loop monomers. Each P-loop is composed of five amino acids: 
T (Threonine, Thr75), V (Valine, Val76), G (Glycine, Gly77), Y (Tyrosine, Tyr78), G (Glycine, Gly79) linked by peptide units (H-N-C=O) in which N-C=O is an amide group and C=O is a carbonyl group. Carbonyls are responsible for trapping and displacement of the ions in the filter  (see Fig.~(\ref{fig2})). At the atomic scale the filter region exposes negative charges owing to the lone electron pairs from oxygen ions bound to 20 carbonyl groups arranged into 5 rings of the lining ’P-loop’ peptide. Alltogether this structure provides a highly ordered atomic coordination pattern among oxygens and ions. If a positive charged ion, such as sodium or potassium enters the selectivity filter, the ion can be transiently trapped by Coulombic interactions with the negative charges provided by the oxygen ions.  

\section{Quantum Superposition States in Ion Channel}
We can expect that the emergence of classicality will involve some quantum signatures in biological systems that cannot be ignored in functional explanations. Answers along this way will most probably play an increasing role for the understanding of organisational complexity and functions in living systems. 
 If quantum effects do play a critical role in filter-ion coordination, it is feasible that these delicate interactions could leave their quantum traces in the overall conformation and the molecular gating state of the entire protein \cite{salari}. The selectivity filter accounts for roughly $1/6$ of the total trans-membrane length of the channel protein. Congruent with the crystallographic $K^+$ electron density profiles, adjacent $K^+$ locations are separated by 0.3 nm in their S1-S3 and S2-S4 configurations \cite{salari}. From a quantum mechanical point of view, these separations can be seen as quantum superposition states of ions with 0.3nm distance in the selectivity filter. It is recently suggested that quantum effects in the selectivity filter can be a reason of highly efficient functioning of ion channels \cite{Vaziri, ganim}. 
 
\begin{figure*}   
   \subfigure[\label{fig31}Tegmark's model in which two ions are in a superposition state of "inside" and "outside" of the membrane. The superposition distance is the thickness of the membrane. Tegmark has assumed 10 nm for the thickness and therefore the superposition distance.]{\includegraphics[width=7cm,height=3cm,angle=0]{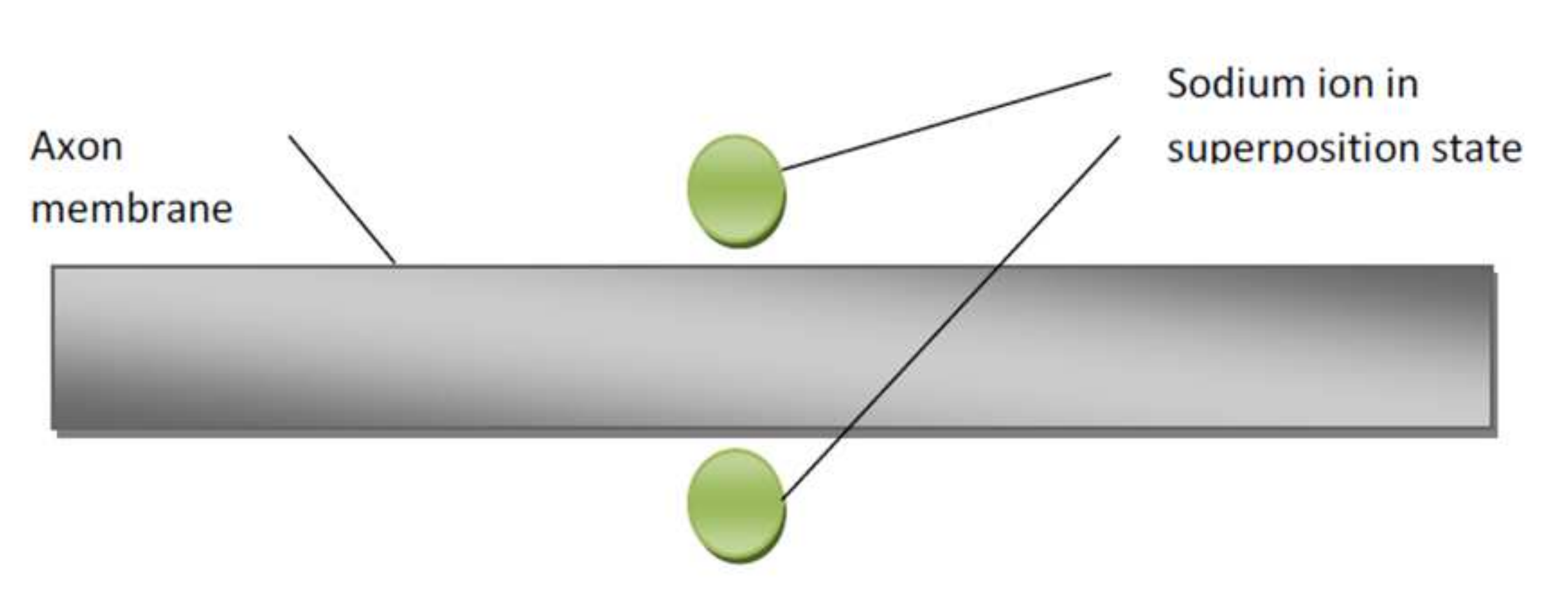}  }~  
       \subfigure[\label{fig3}The Real Model in which quantum superposition states happen in the selectivity filter of ion channel. The superposition distance here is 0.3 nm. The $3.4$ nm long $KcsA$ channel is comprised of a $1.2$ nm long selectivity filter that is composed of four P-loop monomers. Each P-loop is composed of five amino acids linked by peptide units (H-N-C=O) in which N-C=O is an amide group and C=O is a carbonyl group. If a positive charged ion, such as sodium or potassium enters the selectivity filter, the ion can be transiently trapped by Coulombic interactions with the negative charges provided by the oxygen ions.] {\includegraphics[width=8cm,height=6cm,angle=0]{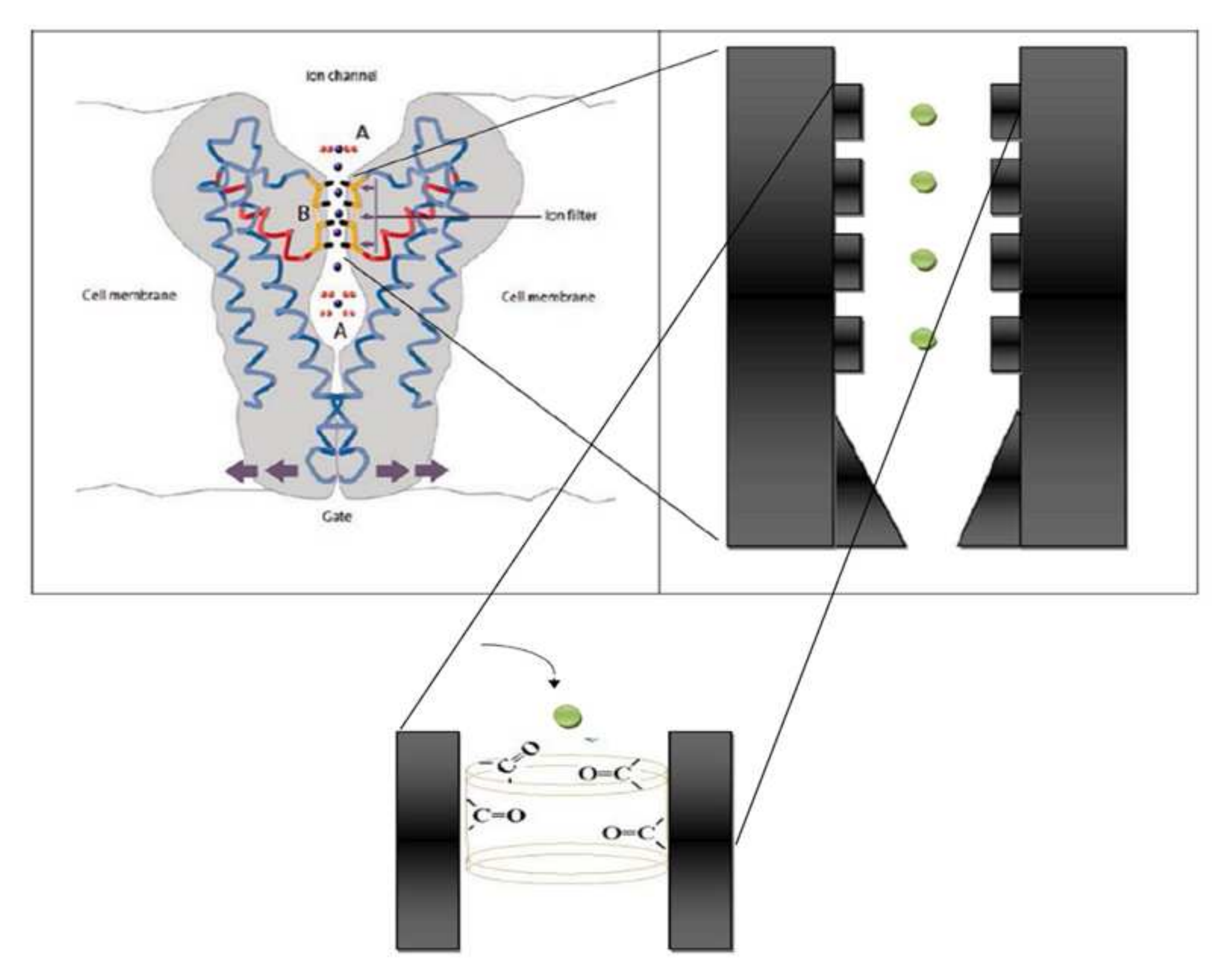}  }
       \caption{The schematic difference between the Tegmark's model and the real model, in which Tegmark has not considered the ion channel structure.}
   \end{figure*}

\section{On the Basis of Decoherence Times Calculation}
Tegmark's calculations to obtain decoherence times \cite{Tegmark2} are based on the scattering of environmental particles from the system. In fact, scattering changes the density matrix of the system, mathematically via multiplying to a function which is the Fourier transform of probability function of the transformed momentum to the system:
   \begin{equation}\label{sh1}
\rho_f(\bm{x},\bm{x}')=\rho_i (\bm{x},\bm{x}')\hat{P}(\bm{x}-\bm{x}')
\end{equation}
where $\rho$ is the density matrix in position basis and $\hat{P}$ refers to the Fourier transform. Then, the temporal distribution of scattering event can be modelled by Poisson distribution with intensity $\Lambda=\sigma\phi$ where $\sigma$ is cross section and $\phi$ is flux, so by this assumption, the evolution of density matrix can be written as \cite{Tegmark1} 
\begin{align}\label{sh2}
\rho(\bm{x},\bm{x}',t+\mathrm{d}t)&=\underbrace{\rho(\bm{x},\bm{x}')\hat{P}(\bm{x}-\bm{x}',t)\Lambda\mathrm{d}t}_{\text{effect of intracted particles}}+\nonumber\\
&\underbrace{\rho(\bm{x},\bm{x}')\hat{P}(\bm{x}-\bm{x}',t)(1-\Lambda\mathrm{d}t)}_{\text{effect of not intracted particles}}\
\end{align}
 arraging this equation leads to
\begin{equation}\label{sh3}
\frac{\partial\rho(\bm{x},\bm{x}',t)}{\partial t}=-\rho(\bm{x},\bm{x}')\underbrace{\left(1-\hat{P}(\bm{x}-\bm{x}',t)\right)}_{F(\bm{x}-\bm{x}')}
\end{equation}
above equation has simple solution. We can define decoherence time as the time in which the $\rho$ reaches to $e^{-1}$ time of its initial value, so
 \begin{equation}\label{sh4}
 \tau_{dec}=\frac{1}{F(\bm{x},\bm{x}')}
 \end{equation}
 in order to calculate $\hat{P}$ we can write it's Cumulant expansion as 
 \begin{equation}\label{sh5}
\ln \hat{P}(\bm{r})=\frac{-i}{\hbar} \la Q_j \ra_c r_j-\frac{1}{2\hbar^2}\la Q_i Q_j \ra_c r_i r_j+\mathcal{O}(x^3)
\end{equation}
where $\la Q_i\ra_c$ and $\la Q_i Q_j\ra_c$ are respectively the first and second Cumulant of distribution.  if we consider the isotropic distribution, we have simply $\la \bm{Q} \ra_c=\la \bm{Q} \ra=0$ and the covariance matrix is proportional to identity matrix $\la Q_i Q_j \ra_c=s^2 \delta_{ij}$, and thus we obtain 
\begin{equation}\label{sh6}
\hat{P}(\bm{x}-\bm{x}')=\exp\left(\frac{s^2|\bm{x}-\bm{x}'|^2}{2\hbar^2}\right)
\end{equation}
and consequently the decoherence time is determined as follows
\begin{equation}\label{sh7}
\tau_{dec}=\frac{1}{\Lambda\left(1-\exp\left(\frac{s^2 \left|\Delta \bm{x}\right|^2}{2\hbar^2}\right)\right)}
 \end{equation}
 we can now define a characteristic length by using  de Broglie relation as $\lambda_{eff}=\hbar/s$ \cite{Tegmark1}. In scattering approach, decoherence time can be expressed in two limits, long and short wavelengths \cite{schloss}:
\begin{equation}\tau_{dec}=\left\{
\begin{array}{ll}
      \frac{1}{\Lambda} &  \lambda \ll \left|\Delta \bm{x}\right| \\
      \frac{\lambda^2}{\Lambda \left|\Delta \bm{x}\right|^2} & \lambda \gg\left|\Delta \bm{x}\right|
\end{array}\right.
\end{equation}
In the short wavelength limit, $\lambda \ll\left|\Delta \bm{x}\right|$, we expect that the particle will be able to well resolve this separation $\left|\Delta \bm{x}\right|$ and thus carry away a maximum of which-path information, inducing a maximum amount of decoherence in the system per scattering event. In the opposite limit, i.e. long wavelength $\lambda \gg \left|\Delta \bm{x}\right|$, the scattered particle will not be able to resolve the separation $\left|\Delta \bm{x}\right|$, and it will thus carry away an only insufficient amount of which-path information. Therefore, we anticipate that it will take a large number of scattering events to induce a significant degree of spatial localization of the object \cite{schloss}.\\
Considering Tegmark's calculations for the system of neuron\cite{Tegmark2}, he has assumed a quantum superposition of  $‘‘resting’’$ and $‘‘firing’’$ states for an order of a million ions being in a spatial superposition of inside and outside the axon membrane, separated by a distance about $h\sim10 nm$ (see Figure 2).
At the room temperature, the de-Broglie wavelength of $Na^+$ is about $\lambda \approx 0.03 nm$ and their spatial separation is $h\sim 10 nm$ (which is assumed to be superposition distance in coherent state). So, based on Tegmark's assumptions, the calculations should be done in the short wavelength limit $\lambda \ll\left|\Delta \bm{x}\right|$ and thus the decoherence times are obtained in the order of $10^{-20}-10^{-19} sec$ \cite{Tegmark2}. Tegmark has considered classical interactions and therefore has used classical data to obtain decoherence times, however the data are determined via some estimates and approximations that make the results inaccurate. In order to obtain accurate data and using them in the similar method of Tegmark, we use classical molecular dynamics simulation to obtain the speed of scatterers.

\section{Method}
Our molecular dynamics (MD) simulations are based on a model of the KcsA channel (Protein Data Bank, 1K4C.pdb), embedded in a palmitoyloleoyl phosphatidylcholine (POPC) lipid bilayer. The system was built from a cubic box of a 7.8 nm side that comprises KcsA (four subunits of 97 amino acids, 5292 atoms), water molecules (TIP3P model, 42296 atoms), 3K and 2K in the pore and 12 CL in the bulk (the entire system is electrically neutral). The AMBER 03 force field parameters \cite{Cornell} and GROMACS 4.5.3 software \cite{Hess} was employed to perform the simulations with the time step of 1 fs. The protein was equilibrated during 10 ps in (N, V, T) then (N, P, T) ensembles. The temperature was kept at 300 K by Nose-Hoover coupling algorithm and the pressure was kept at 1 bar by Parrinello-Rahman coupling algorithm. The system is oriented along the z-axis. A cutoff was used for long-range interactions, namely: 0.12 nm forthe van der Waals interaction and 0.14 nm for electrostatic interactions. Using  the Particle-Mesh Ewald  (PME)  method,  the  electrostatic  interactions  are  calculated. We have used the following abbreviations for amino acid identification: GLY1=GLY79, TYR=TYR78, GLY2=GLY77, VAL=VAL76, THR2=THR75, THR1=THR75 (i.e. hydroxyl). Each carbonyl group is a C=O compound in which the vibrations of C and O atoms are investigated separately.

\begin{table*}[t]\label{tab2}
\caption{The obtained average velocities and wavelenghts of scatterers via MD simulation}
\begin{ruledtabular}
\begin{tabular}{|c|c|c|c|c|c|c|c|c|}
\hline 
 Scatterer & \multicolumn{4}{c|}{Average velocity, MD simulation  (m/s)} & \multicolumn{4}{c|}{Wavelenghts of particles,$\lambda=\dfrac{h}{mV} (nm)$} \\ 
\hline 
 & $V=30mV$ & $V=100mV$ & $V=-70mV$ & $V=-100mV$ & $V=30mV$ & $V=100mV$ & $V=-70mV$ & $V=-100mV$ \\ 
\hline 
K& 194.306 & 192.455 & 293.331 & 300.113 & 0.052 & 0.053 & 0.048 & 0.039 \\ 
\hline 
GLY79.C & 278.748 & 352.612 & 387.051 & 360.738 & 0.111 & 0.094 & 0.086 & 0.092 \\ 
\hline 
TYR78.C & 304.162 & 389.912 & 349.021 & 357.982 & 0.109 & 0.085 & 0.095 & 0.093 \\ 
\hline 
GLY77.C & 368.670 & 354.707 & 355.856 & 347.748 & 0.090 & 0.093 & 0.093 & 0.095 \\ 
\hline 
VAL76.C& 356.809 & 330.769 & 357.842 & 358.014 & 0.093 & 0.100 & 0.093 & 0.093\\ 
\hline 
THR75.C & 355.118 & 384.327 & 340.801 & 382.034 & 0.093 & 0.086 & 0.097 & 0.087 \\ 
\hline 
THR74.C & 330.114 & 368.278 & 340.526 & 359.300 & 0.095 & 0.094 & 0.097 & 0.092 \\ 
\hline 
GLY79.O & 305.631 & 330.934 & 319.312 & 335.160 & 0.081 &0.075 & 0.078 & 0.074 \\ 
\hline 
TYR78.O & 297.581 & 329.592 & 291.042 & 314.781 & 0.084 & 0.076 & 0.085 & 0.079 \\ 
\hline 
GLY77.O& 340.031 & 339.197 & 301.544 & 322.577 & 0.073 & 0.073 & 0.082 & 0.077 \\ 
\hline 
VAL76.O & 304.415 & 313.096 & 296.747 & 306.461 & 0.082 & 0.079 &0.084 & 0.081 \\ 
\hline 
THR75.O & 318.083 & 322.606 & 331.024 & 301.354 & 0.078 & 0.075 & 0.075 & 0.083 \\ 
\hline 
THR74.O& 298.019 & 316.435 & 305.402 & 332.251 & 0.083 & 0.079& 0.081 & 0.075 \\ 
\hline 
H2O& 260.572 & 294.317 & 295.635 & 948.372 & 0.085 & 0.075 & 0.075 & 0.023 \\ 
\hline 
\end{tabular} 
\end{ruledtabular}
 
\end{table*}

\begin{figure*} \centering    
       \includegraphics[width=10.5cm,height=8cm,angle=0]{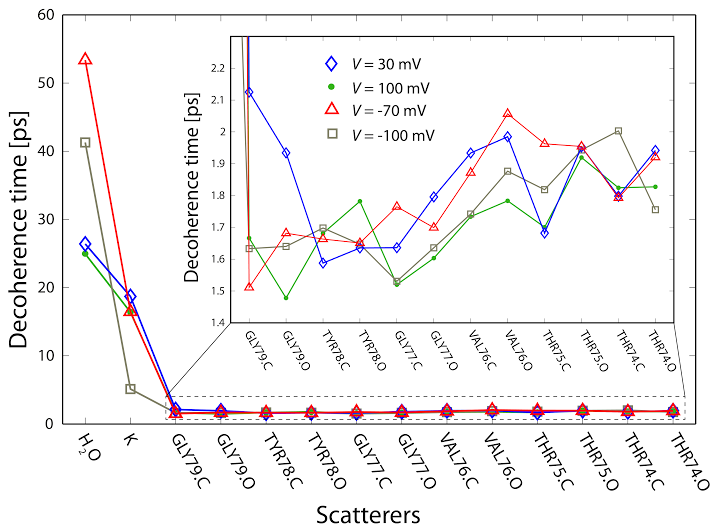}   
       \caption{ Decoherence time of ionic superposition states vs scatterers in the selectivity filter of an ion channel. The results indicate that the main scatterers are carbonyls of amino acids yielding 1ps decoherence time for superposition states.}
        \label{fig3}   
   \end{figure*}

\begin{table*}[t]\label{tab1}
\caption{Comparison of Tegmark Calculations with Our Results}
\begin{ruledtabular}
\begin{tabular}{ccccccc}
 Mechanism &Tegmark& Here & n & N&Tegmark's result&our result\\
 &\scriptsize(Coulombian Interactions)&\scriptsize(MD simulations)&\scriptsize(density of scatterers, Tegmark)&\scriptsize(number of scatterers, Here)&($\approx$) & ($\approx$)\\
 \hline
 Ion-Ion & $\frac{\sqrt{m(k T)^3}}{N g^2 q_e^2 n}$ &  $  \frac{V}{N \sigma  v} $  &  $\frac{10^{23}}{cm^{3}}$ & 2 & $ 10^{-20}s$&$10^{-12}s$\\
Ion-Water & $\frac{\sqrt{m k T}}{N g^2 q_e p n}$ & $  \frac{V}{N \sigma  v} $  & $ \frac{10^{23}}{cm^{3}}$ & 2&$10^{-20}s $&$10^{-12}s$  \\
Ion - Distant ions& $\frac{\sqrt{m k T}}{N g^2 q_e^2 n h}$ & ---- &? &---- &$10^{-19}s$ &---- \\
Ion-Carbonyl \\
(Oxygen) & ------& $  \frac{V}{N \sigma  v} $& ------ & 20 &----& $10^{-12}s$\\
Ion-Carbonyl \\
(Carbon) & ------& $  \frac{V}{N \sigma  v} $& ------ & 20&----& $10^{-12}s$
\end{tabular}
\end{ruledtabular}
\raggedright m is the mass of the ion,    k is Boltzmann s constant,    T is temperature,    N is the number of scatterers,  p is the electric dipole moment of water molecule, g  is the Coulomb constant, q is the charge of an electron, n is the density of scatterers, $\sigma$ is the cross section and v is the velocity. 
\end{table*}

\section{Results}

  Now, we would like to see how fast will an ionic superposition becomes decohered as a consequence of environmental scattering. Briefly, we must calculate decoherence time, i.e. $\tau=\frac{1}{\Lambda}$ where $\Lambda$ is the rate of interaction. So, to obtain the decoherence time, we need the rate of interaction which is equal to $\Lambda = n \sigma v$, where $n$ is the density of scatterers, $\sigma$ the scattering cross section, and $v$ is the velocity of scatterer \cite{schloss, Tegmark2}. Moreover, we shall remember two limiting cases:  The short-wavelength limit in which each scattered particle completely resolves the separation, and the long-wavelength limit in which many such scattering events are required to decohere the superposition state. Basically, we can simply determine the density and cross section of the scatterers in the filter but we do not know their velocities.  Thanks to MD simulations, we have obtained the average velocities of scatterers which are mainly carbon and oxygen atoms of each amino acid besides the ions and water molecules inside the selectivity filter. The results are shown in the Table 2 for four potentials of the membrane. The range of velocities indicate that we must again use the short-wavelength limit. We have plotted decoherence times of ionic superposition states versus scatterers inside the selectivity filter of an ion channel for different potentials in the Figure 3. Our results indicate that the order of decoherence times are $10^{-12}sec$ (i.e. 1ps) which are $10^7-10^8$ times bigger than the calculated decoherence times by Tegmark \cite{Tegmark2}.

%%%%%%%%%%%%%%%%%%%%%%

%%%%%%%%%%%%%%%%%%%%%%

\section{Conclusion}
Neurons are the key building blocks of the brain information processing system, and ion channels are the main parts of neurons for producing action potentials (or electrical signals) in the brain. It was already shown by Tegmark that the nature of information processing in brain cannot be quantum mechanical because neural superposition states cannot survive for the time order of cognitive functions. Indeed, Tegmark has calculated decoherence times for the systems of "ions" and "microtubules" in neurons of the brain while he did not consider ion channels. Here, we re-investigated decoherence times for ionic superposition states via using the data obtained by molecular dynamics simulations. Our results for superposition states of potassium ions indicate that decoherence times are in the order of picoseconds which are 10-100 million times bigger than the orders calculated by Tegmark (see Table 3). This decoherence time is still not enough for cognitive processing in the brain which are in the order of $10^{-3}-1$sec \cite{Tegmark2}, however the obtained decoherence time for quantum states (e.g. cooled ions \cite{salari}) looks enough to leave their quantum traces on the whole structure of ion channel. As a result, action potentials in neurons of the brain can be affected by quantum effects in ion channels and consequently we cannot make sure that action potential in neuron has not any quantum root, and temporal development of quantum states in channel proteins can propagate into classical ion channel conformations that determine the electrical signal properties of neuronal membranes \cite{salari}. Because the filter states must be correlated with the pore domain gating state of the channel in order to facilitate the co-occurrence of conduction and access of ions to the channel protein, the pore gating states of the channels can be interpreted as "classical witness states" \cite{vedral} of an underlying quantum process in the brain.

%{\em Acknowledgements:-} 

\end{document}